\documentclass[12pt]{article}
\usepackage{epsfig,amsmath,amssymb}

\def\nbslash{\rlap{\hspace{0.02cm}/}{\bar n}}

\def\beq{\begin{equation}}
\def\eeq#1{\label{#1}\end{equation}}
\def\eeqn{\end{equation}}

\def\beqa{\begin{eqnarray}}
\def\eeqa#1{\label{#1}\end{eqnarray}}
\def\eeqan{\end{eqnarray}}

\let\bar=\overbar

\def\Dslash{\not{\hbox{\kern-4pt $D$}}}
\def\dslash{\not{\hbox{\kern-2pt $\del$}}}

\def\msb{{\bar{\ssstyle M \kern -1pt S}}}

\def\Title#1{\begin{center} {\Large {\bf #1} } \end{center}}

\begin{document}
\begin{flushright}
WSU-HEP-1203\\
December 18, 2012
\end{flushright}

\Title{Theory of Inclusive Radiative B Decays:\\ 2012 Update}

\bigskip\bigskip

\begin{raggedright}  

{\it Gil Paz\index{Paz, G.}\\
Department of Physics and Astronomy\\
Wayne State University\\
Detroit, MI 48201, USA}
\bigskip\bigskip
\end{raggedright}

\begin{abstract}
This talk discusses recent developments in the theory of inclusive radiative B decays, focusing mainly on non-perturbative aspects of the CP asymmetry in $\bar B\to X_s \gamma$.
\end{abstract}

\section{Introduction}

\let\thefootnote\relax\footnote{
Proceedings of CKM 2012, the 7th International Workshop on the CKM Unitarity Triangle, University of Cincinnati, USA, 28 September - 2 October 2012.}

This talk discusses recent developments in the theory of inclusive radiative B decays, where recent is defined as being after the CKM 2010 workshop. We focus mainly on the CP asymmetry in $\bar B\to X_s \gamma$, since this is arguably the most important new development in the field.

Radiative B decays in general  and $\bar B\to X_s \gamma$ in particular are important probes of new physics. As such they have received considerable attention both from experiment and theory.  The structure of $\bar B\to X_s \gamma$ decay is different from other inclusive B decays such as $\bar B\to X_{u,c} l \bar\nu$ in that it involves more than just one operator. To describe the decay we need not just the ``direct" decay operator  $Q_{7\gamma}=(-e/8\pi^2)m_b \bar s \sigma_{\mu\nu}F^{\mu\nu}(1+\gamma_5)b$, but the full effective Hamiltonian,  
\begin{equation}\label{Heff}
{\cal H}_{\rm eff}=\frac{G_F}{\sqrt{2}}\sum_{q=u,c}\lambda_q\left(\vspace{-2em} C_1 Q_1^q+C_2 Q_2^q+\sum_{i}\vspace{-2em} C_iQ_i+C_{7\gamma}Q_{7\gamma}+C_{8g}Q_{8g}\right)+{\rm h.c.}\,,
\end{equation}
where $\lambda_q=V^*_{qb}V_{qs}$ ($\lambda_q=V^*_{qb}V_{qd}$) for  $\bar B\to X_{s}\,\gamma$  ($\bar B\to X_{d}\,\gamma$) decays. The most important operators, due to their larger Wilson coefficients, are $Q_{7\gamma}$, $Q_1^q=(\bar q{ b})_{V-A}({\bar s} q)_{V-A}$, and $Q_{8g}=(-g/8\pi^2)m_b {\bar s} \sigma_{\mu\nu}G^{\mu\nu}(1+\gamma_5){ b}$. 

Thanks to theoretical advances in recent years, we know that unlike other inclusive B decays, $\bar B\to X_{s}\,\gamma$ integrated rate is described by an expansion in non-local operators suppressed by increasing powers of $\Lambda_{\rm QCD}/m_b$. Only the lowest term in this expansion, as well as the contributions that arise from the interference of $Q_{7\gamma}$ with itself,  can be described in terms of local operators. In particular, the lowest order term in this expansion is given by the partonic rate $\Gamma(b\to s \gamma)$. Including $\Lambda_{\rm QCD}/m_b$ corrections requires contributions that depend on matrix elements of non-local operators. These arise from resolved photon contributions \cite{Lee:2006wn,Benzke:2010js}. Unlike direct photon contributions in which the photon couples to a local operator mediating the weak decay, the resolved photon contributions arise from indirect production of the photon, accompanied by other soft particles.  For a discussion of the resolved photon contributions see the CKM 2010 proceedings \cite{Paz:2010wu}.

The theoretical work in the field since CKM 2010 has focused on improving the theoretical prediction for $\Gamma(b\to s \gamma)$, studying the resolved photon contributions for the CP asymmetry, and higher order non-perturbative corrections  to $\bar B\to X_{s}\,\gamma$. In the following we review each one of them separately.

\section{Recent developments in $\bar B\to X_s \gamma$: \\ Perturbative Aspects}

Since $\Gamma(\bar B\to X_s \gamma)=\Gamma(b\to s \gamma)+{\cal O}\left(\Lambda_{\rm QCD}/m_b\right)$, one major theoretical effort is to complete the NNLO calculation of $\Gamma(b\to s \gamma)$. Shortly after the CKM 2010 workshop, the ${\cal O}(\beta_0\alpha^2_s)$ corrections from $Q_{8g}-Q_{8g}$ were calculated by Ferroglia and Haisch in \cite{Ferroglia:2010xe}. They conclude that the correction to the branching ratio ``amounts to a relative shift of $+0.12\%$" for the cut of $E_\gamma >1.6$ GeV. The results of  \cite{Ferroglia:2010xe} were confirmed by Misiak and Poradzinski in \cite{Misiak:2010tk}, who also calculated the ${\cal O}(\beta_0\alpha^2_s)$ corrections from $Q_{1}-Q_{8g}$ and $Q_{2}-Q_{8g}$. They conclude  that  ``numerical effects of all these quantities on the branching ratio remain within the $\pm 3\%$ perturbative uncertainty estimated in \cite{Misiak:2006zs}". The complete ${\cal O}(\alpha^2_s)$ calculation of $Q_1-Q_{7\gamma}$ and $Q_2-Q_{7\gamma}$  is underway. For a review see \cite{Misiak:2010dz}. As stated in \cite{Misiak:2011bf},  ``The goal of the ongoing perturbative calculations is to make the ${\cal O}(\alpha^2_s)$ uncertainties negligible with respect to the non-perturbative ones."

\section{Recent developments in $\bar B\to X_s \gamma$:  \\ CP asymmetry}
The measured value of the CP asymmetry in $\bar B\to X_s \gamma$ is given in the latest Heavy Flavor Averaging Group report \cite{Amhis:2012bh} by 
\begin{equation}
 {\cal A}_{X_s\gamma} 
   = \frac{\Gamma(\bar B\to X_s\gamma)-\Gamma(B\to X_{\bar s}\gamma)}   
          {\Gamma(\bar B\to X_s\gamma)+\Gamma(B\to X_{\bar s}\gamma)} 
   = - (1.2\pm 2.8)\%\,.
 \end{equation}
In practice, the experiments that have measured the CP asymmetry, namely BaBar, Belle, and CLEO, measure the CP asymmetry with a cut on the photon energy: $E_\gamma\ge E_0$. The cut $E_0$ is between 1.9 and  2.1 GeV, depending on the experiment.  

Theoretically, the CP asymmetry was thought to be of a perturbative origin. In particular, a ``strong" (CP even) phase arises from  loop effects and hence suppressed by $\alpha_s$. A ``weak"  (CP odd) phase arises from complex CKM factors or from complex Wilson coefficients in the effective Hamiltonian. In the Standard Model (SM) the Wilson coefficients are real, so the latter can only arise in extensions of the SM. The perturbative prediction is given by  \cite{Kagan:1998bh,Asatryan:2000kt}
\begin{equation}\label{direct}
\begin{aligned}
   &{\cal A}_{X_s\gamma}^{\rm dir}(E_0)
   = \alpha_s\,\Bigg\{ \frac{40}{81}\,\mbox{Im}\,\frac{C_1}{C_{7\gamma}} 
    - \frac{8z}{9}\,\Big[ v(z) + b(z,\delta) \Big]\,
    \mbox{Im}\bigg[(1+\epsilon_s)\,\frac{C_1}{C_{7\gamma}}\bigg] \\
   &\hspace{-0.5em} - \frac49\,\mbox{Im}\,\frac{C_{8g}}{C_{7\gamma}}+ \frac{8z}{27}\,b(z,\delta)\,
    \frac{\mbox{Im}[(1+\epsilon_s)\,C_1 C_{8g}^*]}{|C_{7\gamma}|^2}
    + \frac{16z}{27}\,\tilde b(z,\delta)\,\bigg|\frac{C_1}{C_{7\gamma}}\bigg|^2\,
    \mbox{Im}\,\epsilon_s \Bigg\} \,,
\end{aligned}
 \end{equation}
where  $\delta=(m_b-2E_0)/m_b$, $z=(m_c/m_b)^2$ and  $\epsilon_s=(V_{ub} V_{us}^*)/(V_{tb} V_{ts}^*)\approx \lambda^2(i\bar\eta-\bar\rho)$. A simpler, and perhaps more transparent expression, is obtained if one takes $m_c^2={\cal O}(m_b\Lambda_{\rm QCD})$ and expand in $z,\delta={\cal O}(\Lambda_{\rm QCD}/m_b)$. In this limit \cite{Benzke:2010tq}
\begin{equation}
   {\cal A}_{X_s\gamma}^{\rm dir}
   = \alpha_s\,\Bigg\{ \frac{40}{81}\,\mbox{Im}\frac{C_1}{C_{7\gamma}}
    - \frac49\,\mbox{Im}\frac{C_{8g}}{C_{7\gamma}} 
  - \frac{40\Lambda_c}{9m_b}\,\mbox{Im}\bigg[(1+\epsilon_s)\,
    \frac{C_1}{C_{7\gamma}}\bigg] 
    + {\cal O}\bigg( \frac{\Lambda_{\rm QCD}^2}{m_b^2} \bigg) \Bigg\} \,,
    \end{equation}
where  $\Lambda_c(m_c,m_b) \approx 0.38\,\mbox{GeV}$. 
In the SM the $C_i$ are real and only the last term contributes. One finds a triple suppression: in $\alpha_s$, $\mbox{Im}(\epsilon_s)\sim\lambda^2\approx0.05$, and $(m_c/m_b)^2\sim\Lambda_{\rm QCD}/m_b$. As a result,  the perturbative theoretical prediction for the SM is an asymmetry of about $0.5\%$ \cite{Soares:1991te,Kagan:1998bh,Ali:1998rr}. A dedicated analysis  \cite{Hurth:2003dk}  finds 
 ${\cal A}_{X_s\gamma}^{\rm SM}=(0.44_{\,-\,0.10}^{\,+\,0.15}\pm 0.03_{\,-\,0.09}^{\,+\,0.19})\%$,
 where the errors are from  $m_c/m_b$, CKM parameters, and scale variation, respectively.  
 
 Compared to the measured value there is ``plenty of room" for new physics. Indeed there were numerous studies of new  physics, see for example the references within \cite{Benzke:2010tq}. From the experimental side, reaching the required precision to observe the CP asymmetry is one of the stated goals of future B factories \cite{Bona:2007qt, Aushev:2010bq}.  
 
 Unfortunately the theoretical prediction described above is not complete, since it does not include resolved photon contributions \cite{Lee:2006wn,Benzke:2010js}.  Are they important for the CP asymmetry? For the CP averaged rate the direct photon contributions are ${\cal O} (1)$ effect, while the resolved photon contributions are suppressed by $\Lambda_{\rm QCD}/m_b$. For the CP asymmetry,  the direct photon contributions are suppressed, so the resolved photon contributions can compete with them. 
 
 Schematically,  the resolved photon contributions to ${\cal A}_{X_s\gamma}$ factorize to sums of convolutions of perturbatively calculable functions  and non-perturbative ``soft" functions. The soft functions are forward matrix elements of non-local operators. Currently they cannot be extracted from data and must be modeled. It can be shown that the soft functions are real  as a result of parity, time reversal, and heavy quark symmetries \cite{Benzke:2010js}. The perturbative functions  arise form uncut propagators and  loops,  and as such they can be complex. In particular they carry strong phases and affect the CP asymmetry.  
 
 At the lowest order in $\alpha_s$ and $\Lambda_{\rm QCD}/m_b$ the resolved photon contributions to the CP asymmetry were calculated in  \cite{Benzke:2010tq}:
 \begin{equation}\label{resolved}
 {\cal A}_{X_s\gamma}^{\rm res}
   = \frac{\pi}{m_b}\,\bigg\{
    \mbox{Im}\bigg[(1+\epsilon_s)\,\frac{C_1}{C_{7\gamma}}\bigg]\,{ \tilde\Lambda_{17}^c}
    - \mbox{Im}\bigg[\epsilon_s\,\frac{C_1}{C_{7\gamma}}\bigg]\,{\tilde\Lambda_{17}^u} + \mbox{Im}\,\frac{C_{8g}}{C_{7\gamma}}\,
    4\pi\alpha_s\,{\tilde\Lambda_{78}^{\bar B}} \bigg\} \,,
\end{equation}
with
\begin{equation}
\begin{aligned}
{ \tilde\Lambda_{17}^u }&= \frac23\,{\ h_{17}(0)}\\
{   \tilde\Lambda_{17}^c }
   &= \frac23 \int_{4m_c^2/m_b}^\infty\!\frac{d\omega_1}{\ \omega_1}\,
  {  f\bigg( \frac{m_c^2}{m_b\,\omega_1} \bigg)}\,{\ h_{17}(\omega_1)} \\
{\    \tilde\Lambda_{78}^{\bar B}}
   &= 2\int_{-\infty}^\infty\!\frac{d\omega_1}{\ \omega_1}
    \left[ {\ h_{78}^{(1)}(\omega_1,\omega_1) - h_{78}^{(1)}(\omega_1,0)} \right]\,, 
\end{aligned}
\end{equation}
where 
$ f(x) = 2x\ln[(1+\sqrt{1-4x})/(1-\sqrt{1-4x})]$.
The soft functions $h_{ij}$ are in light-cone gauge $\bar n\cdot A=0$,
\begin{eqnarray}\label{eqn:h17}
h_{17}(\omega_1,\mu) 
   &=& \int\frac{dr}{2\pi}\,e^{-i\omega_1 r}\,
    \frac{\langle\bar B| \bar h (0)\,
    \nbslash\,i\gamma_\alpha^\perp\bar n_\beta\,
    g G_s^{\alpha\beta} 
    (r\bar n)\,
     h(0) |\bar B\rangle}{2M_B}  \\
 h_{78}^{(1)}(\omega_1,\omega_2,\mu) 
   &=& \int\frac{dr}{2\pi}\,e^{-i\omega_1 r}\!
    \int\frac{du}{2\pi}\,e^{i\omega_2 u} 
    \frac{\langle\bar B| \bar h (0)\,T^A\,
          \nbslash\,
           h(0)\,
          \sum{}_q\,e_q\,\bar q (r\bar n)\,
          \nbslash\,T^A
          q(u\bar n)
          |\bar B\rangle}{2M_B} \nonumber\,.
\end{eqnarray}
The result in (\ref{resolved}) is model-independent. To obtain a numerical result one has to \emph{conservatively} model the soft functions.  The analysis in  \cite{Benzke:2010tq} follows the same modeling as in \cite{Benzke:2010js}. There are three 
non-perturbative parameters that are needed:  $\tilde\Lambda_{17}^u$,  $\tilde\Lambda_{17}^c$, and $\tilde\Lambda_{78}^{\bar B}$. 

To estimate $\tilde\Lambda_{78}^{\bar B}$, Fiertz transformation as well as the Vacuum Insertion Approximation (VIA) are used to express the soft function $h_{78}^{(1)}$ as the square of the B meson light-cone amplitude $\phi_+^B$. In VIA one has 
 $$ 
\tilde\Lambda_{78}^{\bar B}\bigg|_{\rm VIA}=e_{\rm spec}\,\frac{2f_B^2\,M_B}{9}\int_{0}^\infty\,
d\omega_1\,\frac{\left[\phi_+^B(\omega_1,\mu)\right]^2}{\omega_1}\,,
$$
where $e_{\rm spec}$ is electric charge of the spectator quark in units of $e$. Thus $e_{\rm spec}=2/3$  for $B^-$ and $-1/3$ for $\bar B^0$. The integral can be constrained using  \cite{Lee:2005gza}. One finds that in VIA,   $\tilde\Lambda_{78}^{\bar B}\in e_{\rm spec}[17\,\mbox{MeV},\, 190\,\mbox{MeV} ]$.

$\tilde\Lambda_{17}^u$ and   $\tilde\Lambda_{17}^c$ both depend on $h_{17}$. Since it is a soft function, it has support  over a hadronic range. The convolution, on the other hand, starts at $4m_c^2/m_b\approx 1$ GeV, which leads to a small overlap. One finds that   $- 9\,\mbox{MeV} < \tilde\Lambda_{17}^c < + 11\,\mbox{MeV}$ \cite{inprep} . For $\tilde\Lambda_{17}^u$ there is no suppression and one finds that  $- 330\,\mbox{MeV} < \tilde\Lambda_{17}^u < + 525\,\mbox{MeV}
$. The range is not symmetric since the normalization of $h_{17}$  is $2\lambda_2\approx 0.24\, {\rm GeV}^2$ \cite{Benzke:2010js}. This is the same result as one would obtain from naive dimensional analysis.

Including both direct and resolved contributions using $\mu = 2$ GeV as the factorization scale, the \emph{total}  theoretical prediction for the CP asymmetry in the SM is 
$$
   {\cal A}_{X_s\gamma}^{\rm SM}
   \approx \pi\,\bigg|\frac{C_1}{C_{7\gamma}}\bigg|\,\mbox{Im}\,\epsilon_s\,
    \bigg( \frac{\tilde\Lambda_{17}^u-\tilde\Lambda_{17}^c}{m_b}
    + \frac{40\alpha_s}{9\pi}\,\frac{\Lambda_c}{m_b} \bigg) 
   = \left( 1.15\times\frac{\tilde\Lambda_{17}^u-\tilde\Lambda_{17}^c}{300\,\mbox{MeV}}
    + 0.71 \right) \% \,.
$$
Since a slightly lower factorization scale was used for the direct contribution, it is slightly higher than the $0.5\%$ mentioned above.  Using the above estimates for $\tilde\Lambda_{17}^u$ and   $\tilde\Lambda_{17}^c$, one finds that the CP asymmetry in the SM can be in the range $-0.6\%<{\cal A}_{X_s\gamma}^{\rm SM}<2.8\%$ and it is dominated by non-perturbative effects.

Beyond the SM, the Wilson coefficients can be complex. This changes the theoretical prediction to  
\begin{eqnarray}
   \frac{{\cal A}_{X_s\gamma}}{\pi}
   &\approx& \left[ \left( \frac{40}{81} - \frac{40}{9}\,\frac{\Lambda_c}{m_b} \right) 
    \frac{\alpha_s}{\pi} 
    + \frac{\tilde\Lambda_{17}^c}{m_b} \right] \mbox{Im}\,\frac{C_1}{C_{7\gamma}}- \left( \frac{4\alpha_s}{9\pi} - 4\pi\alpha_s\,e_{\rm spec}\,
    \frac{\tilde\Lambda_{78}}{m_b} \right) \mbox{Im}\,\frac{C_{8g}}{C_{7\gamma}} \nonumber\\
   &&- \left( \frac{\tilde\Lambda_{17}^u - \tilde\Lambda_{17}^c}{m_b}
    + \frac{40}{9}\,\frac{\Lambda_c}{m_b}\,\frac{\alpha_s}{\pi} \right)
    \mbox{Im}\bigg(\epsilon_s\,\frac{C_1}{C_{7\gamma}}\bigg)\,. 
\end{eqnarray}
The dependence on $e_{\rm spec}$ implies that the CP asymmetry is different between charged and neutral B's.  In particular one  finds that the \emph{difference} between the CP asymmetries for charged and neutral B's is given by
\begin{equation}
 {\cal A}_{X_s^-\gamma} - {\cal A}_{X_s^0\gamma}
    \approx 4\pi^2\alpha_s\,\frac{\tilde\Lambda_{78}}{m_b}\,
    \mbox{Im}\,\frac{C_{8g}}{C_{7\gamma}} 
   \approx 12\% \times \frac{\tilde\Lambda_{78}}{100\,\mbox{MeV}}\,
     \mbox{Im}\,\frac{C_{8g}}{C_{7\gamma}}\,. 
\end{equation}
The asymmetry difference vanishes in the SM, which makes this observable sensitive to physics beyond the standard model. As was mentioned in Ritchie's talk in this workshop, there is an ongoing BaBar analysis aimed at measuring ${\cal A}_{X_s^-\gamma} - {\cal A}_{X_s^0\gamma}$  \cite{Ritchie}.  

The resolved photon contribution for the CP asymmetry in $\bar B\to X_d \gamma$ are analogous to $\bar B\to X_s \gamma$ with the replacement of $\epsilon_s$ by $\epsilon_d=(V_{ub} V_{ud}^*)/(V_{tb} V_{td}^*)=(\bar\rho-i\bar\eta)/(1-\bar\rho+i\bar\eta)$.  Including the resolved photon contributions the CP asymmetry is in the range $-62\%<{\cal A}_{X_d\gamma}^{\rm SM}<14\%$.  The relation between the resolved photon contributions for  $\bar B\to X_s \gamma$ and  $\bar B\to X_d \gamma$ implies that the untagged CP asymmetry in $\bar B\to X_{s+d} \gamma$ vanishes in the SM up to tiny $U$-spin breaking corrections \cite{Soares:1991te,Kagan:1998bh,Hurth:2001yb} even after including resolved photon effects. Thus it remains a good probe of new physics.

\section{Recent developments in $\bar B\to X_s \gamma$: \\ Non-Perturbative Aspects}

Shortly before CKM 2012,  Kaminski, Misiak, and Poradzinski considered contributions to $\bar B\to X_s \gamma$ from the decay $b\to s  q\bar q \gamma$  \cite{Kaminski:2012eb}. Such decays are suppressed by small CKM factors and/or small Wilson coefficients. Using \emph{on-shell} quarks they find that ``For $E_0 = 1.6$ GeV or higher, the effect does not exceed 0.4\%". A more rigorous QCD-based treatment  using the methods of \cite{Benzke:2010js}, would show that such contributions are suppressed by at least  $\Lambda^2_{\rm QCD}/m_b^2$. The contributions can be expressed in term of four-quark soft functions similar to the ones that appear at order  $\Lambda_{\rm QCD}/m_b$. Considering the non-perturbative uncertainties at order $\Lambda_{\rm QCD}/m_b$, the contribution of $b\to s  q\bar q \gamma$  is not an issue in practice. Still, it is good to check that suppressed effects are indeed small. 

\section{Conclusions}
Radiative B decays in general and $\bar B\to X_s \gamma$  in particular remain a vibrant area of research. Although the field is mature, there are still surprises... For example, since CKM 2010 we have learned that non-perturbative effects, namely the resolved photon contributions, are the dominant effect in the SM prediction for the CP asymmetry in $\bar B\to X_s \gamma$. In particular one finds that $-0.6\%<{\cal A}_{X_s\gamma}^{\rm SM}<2.8\%$, compared to about  0.5\% from perturbative effects alone. The presence of the non-perturbative effects  also implies that  the difference of the CP asymmetry of  charged and neutral B's, ${\cal A}_{X_s^-\gamma} - {\cal A}_{X_s^0\gamma}$,  is a probe of new physics. 

Looking forward,there are still open questions that remain. What are the non-perturbative effects on the $\bar B\to X_s \gamma$ photon spectrum and what are their implications on the extraction of HQET parameters? 
Comparison between theory and experiment relies on extrapolation from the measured values of $E_\gamma \sim 1.9$ GeV to $E_\gamma  > 1.6$ GeV.  How reliable is the extrapolation?  The issue should be revisited. A similar related issue is resummation, i.e. photon energy cut effects in the theoretical calculation. See Gardi's talk in CKM 2008 \cite{Gardi:2008} and CKM 2008 proceedings \cite{Antonelli:2009ws}. What are the resolved photon effects on other radiative B decays, namely $\bar B\to X_sl^+l^-$? Can the non-perturbative error on the total rate be reduced? For example,  the charge asymmetry, $\Delta_{0-}$, that is directly related to the non-perturbative error on $\Gamma(\bar B\to X_s \gamma)$, was only measured by BaBar. It would be very useful to have a similar analysis done by Belle. In short, we have learned a lot concerning radiative B decays, but there is more work to be done.
\section*{Acknowledgments}
I would like to thank Miko\l aj Misiak for useful discussions and Wolfgang Altmannshofer for his comments on the manuscript.

\end{document}